\documentclass[12pt]{article}
\usepackage{epsfig,latexsym}
\begin{document} 
\bibliographystyle{plain} 
\pagestyle{empty} 
\title{\Large \bf Boundary Effects in the One Dimensional Coulomb Gas} 
\author{
D.S. Dean$^\dag$, R.R. Horgan$^*$ and D. Sentenac$^+$ \\\\
$^\dag$CNRS- Laboratoire de Physique Th\'eorique de l'ENS \\
24 rue Lhomond, 75231 Paris CEDEX 05, France.\\\\
$^*$Department of Applied Mathematics and Theoretical Physics \\
University of Cambridge, Silver St., Cambridge, CB3 9EW, U.K.\\\\
$^+$Service de Physique de l'Etat Condens\'e, CEA,\\
Saclay, 91911, Gif--sur--Yvette, CEDEX, France.
}
\maketitle
\begin{abstract}
We use the functional
integral technique of Edwards and Lenard \cite{edle} to solve the
statistical mechanics of a one dimensional Coulomb gas with boundary
interactions leading to surface charging.  The theory
examined is a one dimensional model for a soap film. Finite
size effects and the phenomenon of charge regulation are studied.
We also discuss the pressure of disjunction for such a film. Even in
the absence of boundary potentials we find that the presence of a
surface affects the physics in finite systems. In general we find that
in the presence of a boundary potential the long distance disjoining
pressure is positive but may become negative at closer interplane
separations. This is in accordance with the attractive forces seen at
close separations in colloidal and soap film experiments and with
three dimensional calculations beyond mean field.
Finally our exact
results are compared with the predictions of the corresponding
Poisson-Boltzmann theory which is often used in the context of
colloidal and thin liquid film systems.
\end{abstract}
\vskip 5 truemm
\hfill DAMTP-97-61

\maketitle
\pagenumbering{arabic} 
\def\half{{1\over 2}}
\def\nn{\nonumber}
\pagestyle{plain} 
\def\OO{\Omega}
\def\sech{{\rm sech}}
\def\n{{\newline}}
 \def\la{\langle}
 \def\ra{\rangle}
 \def\aa{\alpha}
 \def\e{\epsilon}
 \def\bk{{\bf k}}
 \def\bkp{{\bf k'}}
 \def\bqp{{\bf q'}}
 \def\bq {{\bf q}}
 \def\EE{\Bbb E}
 \def\EEx{\Bbb E^x}
 \def\EEo{\Bbb E^0}
 \def\LL{\Lambda}
 \def\PP{\Bbb P^o}
 \def\rr{\rho}
 \def\SS{\Sigma}
 \def\ss{\sigma}
 \def\lll{\lambda}
 \def\dd{\delta}
 \def\ww{\omega}
 \def\DD{\Delta}
 \def\DDt{\tilde {\Delta}}
 \def\kr{\kappa\lb \LL\rb}
 \def\PPx{\Bbb P^{x}}
 \def\kk{\kappa}
 \def\tt{\theta}
 \def\bs{\hbox{{\bf s}}}
 \def\bh{\hbox{{\bf h}}}
 \def\lb{\left(}
 \def\rb{\right)}
 \def\prt{\tilde p}
\def\pt{\tilde {\phi}}
 \def\bb{\beta}
 \def\hal{{1\over 2}\nabla ^2}
 \def\bg{{\bf g}}
 \def\bx{{\bf x}}
 \def\bu{{\bf u}}
 \def\by{{\bf y}}
 \def\hag{{1\over 2}\nabla}
 \def\beq{\begin{equation}}
 \def\eeq{\end{equation}}
 \def\bea{\begin{eqnarray}}
 \def\eea{\end{eqnarray}}
 \def\bec{\begin{center}}
 \def\enc{\end{center}}
 \def\bef{\begin{figure}[htb]}
 \def\enf{\end{figure}}
 \def\cosech{\hbox{cosech}}
 \def\ehalf{{\textstyle{e^2\over 2}}}
\section{Introduction}
Up until 1961 the statistical mechanics of the classical 
one dimensional Coulomb gas was an
unsolved problem. At more or less the same time the problem was solved
by Lenard \cite{le} and independently by Prager \cite{pr}.
 A powerful alternative method
of solution using functional integration was subsequently expounded by
Lenard and Edwards \cite{edle}. A good review of this work may be
found in \cite{lima}. It should be  mentioned here that the two dimensional 
Coulomb gas may also be solved exactly at the temperature where 
$e^2/kT = 2$; this exactly soluble case has been investigated in the
electric double layer geometry by Cornu and Jancovici \cite{coja}. 
The problem of electrostatic interactions is one
of profound importance in the theory of colloidal stability and also in
the understanding of thin liquid films. In these problems one
considers the behavior of an electrolytic fluid between two surfaces
which either model the surface of large colloidal particles or the
surfaces of the thin liquid film. The charging mechanism of the
surfaces is usually of a statistical mechanical origin. For example in
soap films made from sodium dodecyl sulphate (SDS), 
the soap anions have hydrocarbon tails which
are hydrophobic and hence have a preference to lie on the surface of
the film \cite{dese1}. 
In colloidal systems chemical reactions may occur between
the colloid particles and the surrounding electrolytic medium again
leading to surface charging. As the two planes are brought together
surface charge regulation occurs. The precise qualitative behavior is
still only understood within the context of mean field
Poisson-Boltzmann type theories \cite{isra,books}, and at a more 
sophisticated level using
the hyper-netted chain approximation (HNC). Rather surprisingly the HNC
theory predicts, in the context of colloidal systems, that the
electrostatic interactions between the planes may become attractive
for small separations \cite{spbe1, spbe2}; this is supported by
calculations of the fluctuations about the mean field solutions
\cite{po}. 
In the  mean field model applied to soap
films charge regulation is predicted \cite{dese1},
 but no attractive component is
seen to appear within the electrostatic interactions. Interestingly the
point at which charge regulation becomes important in the mean field
model for SDS soap films coincides with the range at which collapse to
a Newton Black Film (NBF) occurs \cite{dese1}. 
There is much indirect evidence that the transition from a normal film
to a  NBF is of first order, for
example it is believed to be exothermic and occurs via a nucleation
process where regions of black film expand over the surface of the
film. In this paper we propose to analyze the exactly soluble one
dimensional version of the model proposed in \cite{dese1}. We shall
use the method of \cite{edle} to solve the problem but we shall
highlight the finite size effects appearing in the problem to gain an
understanding of how charge regulation occurs in the model. We shall
compare our exact results with those of mean field theory to ascertain,
at least in one dimension, the accuracy of the traditional
Poisson-Boltzmann mean field approach. 

The paper is arranged as follows. We formulate a form of the soap film
model used in \cite{dese1} in one dimension. The problem is solved
using the functional integral formalism of \cite{edle} and the limit
of bulk systems is rederived for the sake of completeness. We then
analyze the problem in the case of finite films with surface binding
interactions and discuss the nature of the charge regulation and the
stability criterion for the one dimensional film. We then compare the
mean field Poisson-Boltzmann theory with the exact results. Finally we
conclude with a brief comparison between the qualitative behavior
observed in the one dimensional system and that of experiments.

\section{Analysis}
Here we shall summarize the approach of \cite{edle} and apply it to
the system in which we are interested.
The field theory for the system we shall consider is derived from considering
a model consisting of a monovalent soap molecules whose anions are attracted
to the surface of the soap film by the presence of an effective potential
$V(x)$ which acts on them and whose support is localized at the two adjacent
surfaces of the film. In addition one may add an additional monovalent
electrolytic species. In the grand canonical ensemble if the fugacities
for the soap anions/cations and the electrolyte anions/cations are given by
$\mu_s$ and $\mu_e$ respectively, then the partition function is given
by \cite{edle}
 
\begin{eqnarray} 
&Z& =\lb{\rm det}\lb -{d^2\over dx^2}\rb \rb^{1\over 2}
\int d[\psi] \exp\Bigl[ -{1\over 2}\int \lb {d\psi\over dx}\rb^2 dx
 \nonumber \\
&+& \mu_s \int \lb \exp\lb-i\beta^{1\over 2}e\psi - \beta V(x)\rb
+ \exp\lb i\beta^{1\over 2}e\psi\rb \rb dx+ 2\mu_e\int
\cos\lb\beta^{1\over 2}e\psi\rb dx \Bigr] 
\label{eq:FT}
\end{eqnarray}
The derivation of the representation above is quite standard and relies on the
introduction of the Hubbard Stratonovich field $\psi$. In general dimension
the field theory is a form of the Sine-Gordon field theory and is not generally
soluble. For simplicity we shall chose the form of $V(x)$ to be such that
\beq 
\exp(-\beta V(x)) = 1 + \lll^*\delta(x) + \lll^*\delta(x-L),
\label{eq:vx}
\eeq
i.e. the effective surface potential is highly localized about the boundary 
points $0$ and $L$. The $\lll^*$ appearing in (\ref{eq:vx}) is
similar to the adhesivity introduced by Davies \cite{da} in his
analysis of the surface tensions of hydrocarbon solution although the
idea of such a surface active term goes back to Boltzmann. 
For physically realisable soap films $V(x)$ is not
strictly localized  as the effective potential created due to the 
hydrophobic nature of
the soap anion hydrocarbon tails has a support over a region of the length
the tail between the surface and the interior of the film (see
\cite{dese1} for a discussion of the mechanism generating this
potential). However for the
purposes of demonstrating the the essential physics of charge regulation in a 
one dimensional system our choice of $V(x)$ should be adequate. With this 
choice of $V$ we obtain
\begin{eqnarray}
Z &=& \lb{\rm det}\lb -{d^2\over dx^2}\rb \rb^{1\over 2}
\int d[\psi] \exp\Bigl[- {1\over 2}\int \lb {d\psi\over dx}\rb^2 dx
\nonumber \\
&+& 2\mu\int
\cos\lb\beta^{1\over 2}e\psi\rb dx  + \lll\lb \exp(-i\beta^{1\over
2}e\psi(0))+ \exp(-i\beta^{1\over2}e\psi(L))\rb \Bigr],
\end{eqnarray}
with $\mu = \mu_s + \mu_e$ and $\lll = \mu_s \lll^*$. 
Because now the potential $V$ acts only on the end points we may write $Z$ in
path integral notation as,
\begin{eqnarray} 
Z &=& {1\over N}
\int d\psi_0 d\psi_L \int_{\psi_0}^{\psi_L}
 d[\psi] \exp\Bigl[ -{1\over 2}\int \lb {d\psi\over dx}\rb^2 dx \nonumber \\
 &+& 2\mu\int
\cos\lb\beta^{1\over 2}e\psi\rb dx  + \lll\lb \exp(-i\beta^{1\over
2}e\psi(0))+ \exp(-i\beta^{1\over2}e\psi(L))\rb \Bigr].
\end{eqnarray}
The path integral is that for a diffusing particle in a cosine potential (the
normalization coming from the determinant is absorbed in the factor $N$), 
consequently following \cite{edle} we find
\beq 
Z = {1\over N}
\int d\psi_0 d\psi_L P\lb \psi_0,\psi_L | L\rb
 \exp\Bigl[\lll\lb \exp(-i\beta^{1\over
2}e\psi_0)+ \exp(-i\beta^{1\over2}e\psi_L)\rb \Bigr],
\eeq

where $P(\psi_0,\psi | x)$ obeys
 
\beq 
{1\over 2}{\partial^2 P\over {\partial \psi^2}} + 2\mu\cos\lb\beta^{1\over 2}e\psi\rb P = 
{\partial P\over {\partial x}},
\eeq

subject to the initial condition $ P(\psi_0,\psi | 0) = \delta(\psi_0 -\psi)$.
Note here that the boundary terms $\psi_0$ and $\psi_L$ in our path integral 
are free and are integrated over, this is in contrast to the study of 
\cite{edle}, where $\psi_0 = 0$ and $\psi_L$ is left free. This is because 
in the formulation of \cite{edle}, it is assumed that the overall system bulk
plus the film is electroneutral and that the bulk lies to the left of the point
$0$. In our case the bulk may be screened from the film and the most general
boundary conditions are those we have employed.
In order to determine the normalization factor $N$ we note that
when $e = \lll = 0$ then we should obtain the ideal gas result 
$Z = \exp(2\mu L)$. In this case

\beq
 P(\psi_0,\psi | x) = {1\over {(2\pi x)^{1\over 2}}} \exp\lb - 
{1\over 2x}(\psi -\psi_0)^2 \rb \exp(2\mu x)
\eeq
giving simply 

\beq 
Z|_{e =\lll = 0} = {1\over N}\int d\psi_0 \exp(2\mu L).
\eeq

At this point we must regularize the partition function by bounding the 
possible values of $\psi_0$ between two extrema yielding

\beq
N = {1\over {{\rm max}(\psi_0) - {\rm min}(\psi_0)}}.
\eeq

In the case where $e\neq 0$ we may use the fact that the action is
invariant under translations of ${2\pi/ e \beta^{1\over 2}}$
to obtain (we have simply assumed that the extremal values of $\psi_0$ are 
integer multiples of ${2\pi/ e \beta^{1\over 2}}$)

\beq
Z = {e\beta^{1\over 2}\over 2 \pi}\int_0 ^{2\pi/\beta^{1\over 2}}
d\psi_0 \int_{-\infty}^{\infty} d\psi_L  \exp\Bigl[\lll\lb 
\exp(-i\beta^{1\over2}e\psi_0)+ \exp(-i\beta^{1\over2}e\psi_L)\rb\Bigr]
 P\lb \psi_0,\psi_L | L\rb.
\eeq

A further simplification is obtained by noting that

\beq
\int P\lb \psi_0,\psi_L | L\rb f(\psi_L)d\psi_L = K(\psi_0 |L),
\eeq
where $K(\psi_0 | L)$ obeys

\beq
H K = {\partial K \over \partial L},
\eeq

with

\beq 
H = {1\over 2}{\partial^2\over \partial \psi_0^2} + 2\mu 
\cos(\beta^{1\over2}e\psi_0),
\eeq

subject to the initial condition $ K(\psi_0|0) = f(\psi_0)$. In our problem 
\beq
f(\psi_0) =
\exp\lb\lll \exp(-i\beta^{1\over
2}e\psi_0)\rb \label{F}
\eeq

For simplicity in notation we shall take $\beta = 1$ from here on. To recover
the temperature dependence the rescalings $e^2 \rightarrow \beta e^2$ and
$P \rightarrow kTP$ should be performed. The final expression for $Z$ is

\beq 
Z = {e\over 2\pi}\int_0^{2\pi \over e} d\psi_0 K(\psi_0 |L) f(\psi_0),
\eeq

which in operator notation may be expressed as

\beq
 Z = {e\over 2\pi}\int_0^{2\pi \over e} d\psi_0
 f(\psi_0) \exp(LH)  f(\psi_0).
\eeq

The free energy of the film is simply given by $F(L)= -\log(Z)$. In order to
calculate the disjoining pressure we assume that the film is attached to an 
infinite reservoir of particles with the same chemical potentials.
In addition the over all volume (i.e. length) of the system is
conserved. This last point is quite important in understanding the
statistical physics of small systems where thermodynamic ideas cannot
be necessarily applied directly. Our physical assumption of the over
all incompressibility of the system is motivated by the principle that
the background space of water can be assumed to be incompressible 
in most experimental regimes in colloid and thin film science.

Therefore on changing the volume of the film by $\delta L$, the free
energy of the bulk is changed by $-\delta L \theta_\infty$ where 
$\theta_\infty$ is the free energy per unit volume of an infinite system 
(i.e. one which is really in the thermodynamic limit) without any boundary 
interaction. The work done on this change is 
 $P_d \delta L$ where $P_d$ is the disjoining pressure. Therefore
we find that

\beq
 P_d = -{\partial F\over \partial L} + \theta_\infty .\label{P_D}
\eeq

Finally defining the field $\phi = e\beta^{1\over 2}\psi$ we obtain

\beq 
Z = {{1\over 2\pi}}\int_0^{2\pi} d\phi_0 f(\phi_0) \exp(LH)  f(\phi_0),
\eeq

where now $f(\phi_0) = \exp(\lll\exp(-i\phi_0))$ and 

\beq
 H = e^2{1\over 2}{\partial^2\over \partial \phi_0^2} + 2\mu \cos(\phi_0).
\eeq
One point observables are calculated as
\beq \la O(\phi(x))\ra = {1\over Z}
\int_0^{2\pi} d\phi_0 f(\phi_0) \exp(xH)O(\phi_0)\exp((L-x)H)
f(\phi_0),
\eeq
the extension to n-point functions being trivial.
Here for convenience and to highlight the different regimes we
develop the notation. We write
\beq 
H = {e^2\over 2} H^* 
\eeq
where
\beq 
H^* ={ \partial^2\over \partial \phi_0^2} + a\cos(\phi_0) \label{HST}
\eeq
with $a = 4\mu/e^2$. 
\section{\label{larege_films} Results for Large Films: Thermodynamic Limit}
The eigenfunctions of $H$ periodic on $[0,2\pi]$
are the periodic Mathieu functions $\chi_n(\phi,a)
$ whose eigenvalues we denote by
$\gamma_n(a)$ and where it is easy to see that the largest eigenvalue
$\gamma_0(a) < a$. Hence in the case where $e^2 L/2 \ll 1$

\beq
 Z \sim {1\over 2\pi}\lb \int_0^{2\pi}d\phi
\chi_0(\phi)f(\phi)\rb^2 \exp({e^2\over 2}L\gamma_0(a)).
\label{eq:zthick}
\eeq

therefore  $\theta_{\infty} = -{e^2\over 2}\gamma_0(a)$ and
any boundary terms become insignificant in the thermodynamic limit.
The bulk pressure is then $P_{bulk} = -\theta_{\infty}~$.

\subsection{Small $a~$, large $e$}
In the case where $a$ is small one may evaluate $\gamma_0(a)$
in perturbation theory and we find that \cite{abst}

\beq
\gamma_0(a)~=~{1\over 2}a^2~-~{7\over 32}a^4~+~{29\over 144}a^6~-~
              {68687\over 294912}a^8~+~\ldots~.\label{GAMMA0}
\eeq

In this regime the average density is extensive and is given by

\beq
\rho~=~\mu{\partial P\over \partial \mu}~=~
{e^2\over 2}\,a{d\,\gamma_0(a)\over da}~,\label{RHO}
\eeq
The condition that $a$ is small is therefore equivalent to
\beq
a~\sim~\sqrt{2\rho}/e \ll 1~.
\eeq
This implies that the electrostatic energy is much greater than the
contribution from the entropy. From equations (\ref{P_D},~\ref{GAMMA0},
~\ref{RHO}) the pressure is then given by the series
\beq
P_{bulk}=~{1\over 2}\,\rho\left[1~+~{7\over 8}\lb{\rho\over e^2}\rb~-~
{23\over 288}\lb{\rho\over e^2}\rb^2~-~
{4897\over 122288}\lb{\rho\over e^2}\rb^3~+~\ldots\right]~. \label{P_SA}
\eeq
This result was explained by Lenard as an effect of dimerization.
The leading term is independent of $e$ and is the perfect gas result for
a density of $\rho/2$. This is explained by the positive and
negative charges binding in pairs to give, in leading order, a neutral
gas with half the original particle density. The non-leading terms correspond
to multipole interactions such as van der Waals forces etc.

\subsection{Large $a~$, small $e$}

The calculation of $\gamma_0(a)$ for large $a$ can be formulated as
a perturbation series for $H^*$ in equation(\ref{HST}) obtained by
expanding the cosine and writing
\beq
H^*~=~a~+~H_{osc}~+~O(a\,\phi^4)~,\label{HST_OSC}
\eeq
where $H_{osc}$ is the harmonic oscillator Hamiltonian
\beq
H_{osc}~=~{\partial^2\over \partial \phi^2}~-~\half a\,\phi^2~.
\eeq
Thus the perturbation theory is for the anharmonic terms in
equation (\ref{HST_OSC}) using the basis of oscillator states
associated with $H_{osc}~$. The first term in the pressure is
due to the $O(a)$ term in equation (\ref{HST_OSC}) and gives the
free gas contribution. The next correction arises from the
ground state eigenvalue of the oscillator and is $O(\sqrt{a})~$.
This is the well-known Debye-H\"uckel term. We expect a power
series in $a^{-1/2}$, but to carry out the perturbative expansion
becomes increasingly difficult as the order increases. Instead, we
can formulate the problem as an expansion in Feynman diagrams.
A similar approach to the calculation of the electrostatic free
energy of a system of fixed charge macroions has been used
by Coalson and Duncan \cite{codu} and Ben-Tai and Coalson \cite{btco}.
In the bulk limit we write the Feynman kernel for $\half H^*$ as
\beq
{\cal K}(L,a)~=~N\int d\psi\,\exp\lb \int_{-L/2}^{L/2}dx\,
\left[-\half\lb{d\psi\over dx}\rb^2~+~{a\over 2}\,\cos(\psi)\right]\rb~,
\label{K}
\eeq
where $N$ is the normalization factor chosen so that ${\cal K}(L,a=0) = 1~$.
In the limit $L \rightarrow \infty$
\beq
{\cal K}(L,a)~=~\exp({\gamma_0(a)\over 2}L)~~\Rightarrow~~\gamma_0(a)~=~
                                          2{\partial\over \partial L}\log{
{\cal K}(L,a)}~.
\eeq
It is convenient to discretize $L$ so that $L=n\e~$, where $\e$ is
the lattice spacing. Then the operation $\partial/\partial L$ can be performed directly
on the LHS of equation (\ref{K}) as
\bea
\gamma_0(a)~&=&~2{1\over n}{\partial\over\partial\e}\,\log{{{\cal K}(L,a)}}\nn\\
&=&~\la\lb{\psi^\prime}\rb^2\ra~+~
                 a\la\cos(\psi)\ra~-~{1\over \e}~.\label{FD}
\eea
The last term arises because the normalization factor $N$ depends on $\e~$.
This term cancels trivially with a simple $\e$ divergence in the 1-loop graph
for $\la (\psi^\prime)^2\ra~$. We can now take the limit $\e \rightarrow 0~$.
The density, defined by equation (\ref{RHO}), is given by
\beq
\rho~=~{e^2a\over 2}\la\cos(\psi)\ra~,\label{RHOCOS}
\eeq
and thus the second term in equation (\ref{FD}) corresponds to the free gas term. Hence,
we have
\beq
P_{bulk}~=~\rho~+~{e^2\over 2}\left[\la\lb\psi^\prime\rb^2\ra~-~{1\over \e}\right]~.\label{P_FD}
\eeq
We define the Debye mass $m$ by $m^2 = a/2$ and then
\beq
{\cal K}(L,a)~=~\int d\psi\,\exp(-S(\psi))~,
\eeq
where overall irrelevant constant factors have been omitted, and
\beq
S(\psi)~=~\int_{-\infty}^\infty dx\,\left[\half\lb\psi^\prime\rb^2~+~\half m^2\psi^2~+~
              {(-1)^{n+1}m^2\over (2n)!}\psi^{2n}\right]~.
\eeq

A standard Feynman graph expansion of closed loops for $\la\lb\psi^\prime\rb^2\ra$
can be obtained and hence the pressure can be calculated from equation (\ref{P_FD}).
Standard dimensional analysis shows that the series obtained is in inverse powers of
$m$ and that a diagram with $l$ loops behaves as $m^{2-l}~$. To three-loop order
we evaluate the diagrams shown in figure \ref{f1} and find
\beq
P_{bulk}~=~\rho~+~{e^2\over 2}\lb -{m\over 2}~+~{1\over 16}~+~{3\over 512m}\rb~\ldots~.
\label{P_LA}
\eeq
In order to express $P_{bulk}$ as a function of the density, $\rho$, we use equation (\ref{RHOCOS})
and calculate $\rho$ in the loop expansion. As before, the diagrams with $l$ loops
behave as $m^{2-l}~$. If $P$ is calculated to $l$-loop order, then $\rho$ is needed to $(l-1)$-loop
order. To two-loop order we find
\beq
\rho~=~e^2m^2\lb 1 - {1\over 4m}\rb~.\label{RHOEXP}
\eeq
Note that, alternatively, 
\beq
\rho~=~{m\over 2}\,{\partial P_{bulk}\over \partial m}~,
\eeq
and if we define $t=\log(m)$ and write $~P_{bulk} = \rho + P_e~$ then we have
\beq
{d\rho\over dt} - 2\rho~=~-{dP_e\over dt}~,
\eeq
which has solution
\beq
{\rho\over m^2}~=~\int_m^\infty \,{1\over m^2}\:{dP_e\over dm}\:dm~+~1~,
\eeq
where we have used the boundary condition that $\rho/m^2 \rightarrow 1$ as $m\rightarrow \infty~$.

Then $P(m)$ can be re-expressed as a series in $\sqrt{\rho/e^2}~$:
\beq
P_{bulk}~=~\rho~-~{1\over 4}\sqrt{\rho e^2}~+~{1\over 1024}\sqrt{e^6\over\rho}~+~\ldots~.\label{P_LA0}
\eeq
This agrees with Lenard \cite{le} and it is relatively easy to evaluate the
loop expansion to higher orders to improve on Lenard's result. The second
term is the familiar Debye-H\"uckel contribution and it should also be noted that
the two-loop contribution is zero.

\subsection{The bulk pressure}
In figure \ref{f11} we show the computed value for $P_{bulk}$ compared with
the predictions of the previous two sections. For convenience $P_{bulk}$
has been scaled by a factor of $e^2/2$. As can be seen the curves from
equations (\ref{GAMMA0} and (\ref{P_LA}) fit very well except in the
region $0.7 < a < 1.1~$ where even so the discrepancy is not very large.

\section{Thin Films: Finite size Effects and Surface Charge Regulation}

When the intersurface distance of the film becomes small in the sense
that ${1\over 2}e^2 L$ is no longer large, then we may not apply the
thermodynamic result (\ref{eq:zthick}). 

However if ${1\over 2}a e^2 L$ and ${1\over 2}e^2 L$ are both small,
which is certainly the case for extremely small $L$, then one may
expand the operator $\exp(LH)$ in powers of $LH$. To second order in
$LH$ one obtains
\beq Z = 1 + 2\mu L\lll + {L^2\over 2}(2\mu^2(\lll^2 + 1) -\mu\lll
e^2) + O(L^3)\eeq
thus yielding $F \approx -2\mu L\lll$ and  $P_d =2 \mu\lll +
\theta_\infty$. Hence one has the limiting value of the disjoining
pressure is negative if the value of $\lll$ is sufficiently
small. However, one has the bound that $\gamma_0 (a) < a$ and hence 
$\theta_{\infty} > - 2\mu $ thus
\beq \lim_{L\to 0} P_d > 2\mu  \lll - 2\mu .\eeq
Hence the film certainly  has a positive disjoining pressure  at small
differences if $\lll >1$.
The stability of the film at small separations is determined by
\beq 
\lim_{L\to 0} {\partial P_d\over \partial L} = 2\mu^2(1-\lll^2) -\mu
\lll e^2. 
\eeq
If this term is positive then the film collapses to the point
thickness $L = 0$. For this to happen one must have
\beq \lll < 1, \eeq
and
\beq \mu > {\lll e^2 \over 2(1-\lll^2)}. \eeq

The
value of the surface charge $\sigma$ is given by
\beq 
\sigma = -\lll e \la \exp(-i\phi(0)) \ra =
-eL\lll\mu~.
\label{eq:sigex} 
\eeq
Hence over short distances the surface charge decays linearly as the
two surfaces are brought together. 
 
\section{\label{INTER} Intermediate Regime}
In the regime between very thick and very thin films we shall resort
to a numerical analysis of the problem. There are two methods of
interest which we detail in the following sections. For convenience
of notation we shall work in units where $L$ is scaled by $e^2/2$.
That is, a factor of $e^2/2$ is absorbed into all length variables.
\subsection{\label{M_sec} The Mathieu function method}                             
The disjoining pressure and other properties of the film can be calculated
using the even and odd Mathieu functions that are the eigenfunctions
of $H^*$ defined in equation (\ref{HST}). The kernel ${\cal K}(L,a)$ defined
in equation (\ref{K}) can be computed as an expansion on the Mathieu 
function by resolution of the identity on the basis of these states.
In this way the disjoining pressure may be in general be written as               
\beq                                                                
P_d(L) = -{e^2\over 2}{\sum_{n=1}^\infty\;(\gamma_0(a) -         
\gamma_n(a))\;c_n^2\;\exp(\gamma_n(a)L) \over            
\sum_{n=0}^\infty\;c_n^2\;\exp(\gamma_n(a)L) }, \label{P_MAT}     
\eeq                                             
where
\beq                                                                   
c_n = \int_0^{2\pi} d\phi\,\chi_n(\phi)\exp(\lll\exp(-i\phi)) .         
\eeq                                   
If the eigenvalues of $H$ are arranged in descending order i.e. 
$\gamma_0 > \gamma_1 > \gamma_2 ...$, the corresponding eigenfunctions
are even about $\pi$ for $n$ even and odd about $\pi$ for $n$ odd.
Hence $c_n$ is purely real                                       
for $n$ even and purely imaginary for $n$ odd. If $\lll=0$ then         
$c_n = 0$ for $n$ odd and hence $P_d$ is always negative, hence the force
between the two interfaces is always attractive. One sees from the    
above expression that it is the even wave functions which are                
attractive and the odd wave functions which are repulsive (as the
demoninator on the RHS of (\ref{P_MAT}) is $Z$ and hence positive).
At long distances                 
\beq
P_d \sim  {e^2 \over 2} {c_1^2\over c_0^2}\;(\gamma_0 -
\gamma_1)\;\exp(L(\gamma_1 -\gamma_0))~>~0 \label{eq:pdlong}
\eeq
Hence for a non-zero $\lll$ the long distance disjoining pressure
is always positive since $c_1^2 < 0~$. It is clear however that the
disjoining pressure may become negative at smaller values of $L$.

The anion and cation number densities as a function of $x$, the distance through
the film, may also be calculated. Denoting these densities respectively by 
$\rho_+$ and $\rho_-$ we find
\beq
\rho_\pm~=~
\mu\la f^* | \chi_n \ra\,\exp(-\gamma_nx)\,\la \chi_n | e^{\pm i\phi} | 
\chi_m \ra\,\exp(-\gamma_m(L-x))\,\la \chi_m | f \ra~, \label{D_MAT}
\eeq
where
\beq
\la \psi | \chi \ra~=~\int d\phi\,\psi^*(\phi)\chi(\phi)~.
\eeq

We are able to construct both the even and the odd Mathieu 
functions and their eigenvalues for any value of $a$ using Given's method for 
diagonalizing a tri-diagonal matrix. The eigenfunctions of $H^*$ are found on a 
discretization of the interval $[0,2\pi]$ and the appropriate matrix elements in 
equations (\ref{P_MAT}, \ref{D_MAT}) can be calculated numerically.

\subsection{\label{F_sec} Fourier method}
It turns out that there is a more direct method to calculate the disjoining 
pressure which exploits the periodicity inherent in the system. This method
is especially effective for low temperature (small $e$). It does, however, become
much more complicated when other observables such as the density profiles are
being calculated. Expanding $K$ in terms of its Fourier modes,  i.e. writing

\beq 
K(\phi |x) = \exp(ax)\sum_{n =-\infty}^\infty b_n(x)\exp(in\phi),
\eeq
 
one has that

\begin{eqnarray} 
b_n(0) &=& {\lll^{-n} \over (-n)!} \hbox{ for } n \leq 0 \nonumber \\
       &=& 0 \ \ \ \ \ \ \hbox{ for } n > 0
\end{eqnarray}
 
and the $b_n$ evolve via the equation
\beq 
{db_n \over dx} = -n^2 b_n + a(b_{n+1}+ b_{n-1} - 2 b_n)/2.
\eeq
Finally the partition function is given by
\beq 
Z = \exp(aL)\sum_{n =0}^\infty {\lll^{n} \over (n)!}b_n.
\eeq
 
In this regime we shall also be interested in the mean value of the 
surface charge $\sigma$. Including a source term in the original
formulation of the problem it is a simple matter to show that

\beq 
\sigma = \lll e \la \exp(-i\phi)\ra =
{\lll e\over Z}
  {{1\over 2\pi}}\int_0^{2\pi} d\phi_0 \exp(-i\phi_0)f(\phi_0)
 \exp(LH)  f(\phi_0). \label{F_SIG}
\eeq

In terms of the Fourier expansion this becomes

\beq 
\sigma = {1\over Z}\exp(aL)\sum_{n =0}^\infty {\lll^{n} \over
(n)!}b_{-n+1}
\eeq

The disjoining pressure may be computed similarly.

In what follows we shall consider three cases which are paradigms 
for the different regimes of high, intermediate and low temperature.
Since we have set $kT = 1$, high $T$ corresponds to a small charge
parameter, $e$ and vice-versa. Apart from an overall dimension-carrying
factor the results depend on $e$ and $\mu$ through the combination
$a=4\mu kT/e^2$, and in what follows we choose $\mu=1$ and hence in
our units $a=1/e^2~$. The three regimes of temperature are characterized
by the three values of charge: $e = 0.1,1,4~$.

\subsection{$e=0.1,~a=400$}
From equation (\ref{P_LA}) the bulk pressure is $P=1.926$. The major
correction to the free particle pressure, $P_{free}=2$, is the Debye-H\"uckel
term and the two and three-loop contributions are a correction of only
$\Delta P = 0.003~$. In figures \ref{f2} and \ref{f3} we show the
pressure 
$P$ versus film thickness $L~$ for various values of $\lll$ in the
range $0.9$ to $1.2~$. Also plotted is the prediction for the bulk
pressure to which all curves should be asymptotic. As can be seen there
is a collapse in all cases shown for $\lll$. The details of the collapse
differ, however, as $\lll$ increases. For the lower values of $\lll$ the
collapse is to a film of zero thickness which would, of course, be dominated
by the detailed structure of the surface physics which we have subsumed
in to a layer of zero thickness. Two maxima are clearly visible for
$\lll = 0.93, 0.95$. The one at larger $L$ is the location of the
ordinary collapse point. The maximum at smaller $L$ and the 
consequent multiple-valuedness of the curve in $L$ versus $P$ in
this region implies a hysteresis phenomenon as $P$ is cycled
for very thin films. This kind of effect is reminiscent
of a first-order transition which predicts that for a 3D
film there will be domains of different thicknesses which will grow
or contract like 2D bubbles. Of course, it remains to be seen whether
intuition from 1D survives for the realistic 3D case. The typical length 
scale is $O(e^2/2) \sim 5\,10^{-3}$, which is very small compared with
the values of $L$ plotted in figures \ref{f2} and \ref{f3}. 

For the larger values of $\lll$ plotted the curves  the collapse is to 
a thinner film but not to one of zero thickness. as $\lll$ increases the maximum
at small $L$ eventually disappears and for much larger $\lll$ the collapse
phenomenon itself disappears.

In figure \ref{f4} we shown the surface charge $\ss$ defined in equation
(\ref{F_SIG}) as a function of $L$. There is no feature which hints at the
presence of the collapse phenomenon appearing in the associated pressure curves, 
but in all cases $\ss$ decreases with $L$. For small $L$ the behavior agrees
well with the prediction of equation (\ref{eq:sigex}).

The anion and cation densities have been computed as a function of $x$ for various
values of $L$ using equation (\ref{D_MAT}). For $e=0.1$ the variation with $x$ is mild 
and shows no features of note. We show the midplane values for each species as a function 
of $L$ and for various values of $\lll$ in figure \ref{f5}.

It is interesting to note that both methods described in subsections \ref{M_sec} and
\ref{F_sec} were used to calculate the disjoining pressure. However, while 10 fourier
modes were amply sufficient, the number of Mathieu modes needed was 40. In particular
this large number of modes was found necessary to reproduce the secondary collapse 
maxima shown in figure \ref{f2}.

\subsection{$e=1.0,~a=4$}
As in the previous section the pressure $P$ versus $L$ is plotted in figure \ref{f6}
for values of $\lll$ in $0.3 \le \lll \le 0.8$ which span a region of collapse. 
In this case there is just one point of collapse to a film of zero thickness
(in our approximation) and which disappears for $\lll$ between $0.7$ and $0.8$.
From equation (\ref{P_LA}) the bulk pressure is predicted to be $P_{bulk} = 1.3262$. From
the exact calculation we find $P_{bulk} = 1.32584$ which is in good agreement with the
prediction. To guide the eye the computed asymptotic value is shown in figure \ref{f6}.
These pressure curves are well reproduced by the Mathieu function method with as few
as 8 modes. Unlike the case in the previous section $a=4$ is sufficiently small that
the physics is dominated by the lowest-lying Mathieu eigenfunctions. This is mainly
due to the fact that $\lll$ is smaller and so the overlap $\la \chi_n | f \ra$ falls
off more sharply with $n$. This in turn means that the pressure peak occurs for larger
$L$ than in the $a=400$ case. However, the large-$a$ result for the bulk pressure,
equation (\ref{P_LA}, \ref{P_LA0}), still holds very well in this region which means that the
Debye-H\"uckel approximation is good.

The surface charge $\ss$ is plotted in figure \ref{f7} and, as in the previous case,
there are no features associated with the pressure maxima of figure \ref{f6}.

The anion and cation number densities as a function of distance, $x$, through the film 
are shown in figure \ref{f8} for $\lll=0.5~$. These quantities were calculated using
equation (\ref{D_MAT}). The anion (cation) curves are the higher (lower) set in this
figure. There are no unusual features and the curves for the other values of $\lll$
in $0.3 \rightarrow 0.8$ are of similar form.

\subsection{$e=4.0,~a=0.25$}
The pressure $P$ is plotted versus $L$ for $\lll=0.12, 0.121, 0.122, 0.123$ in
figure \ref{f9}. The collapse region is again evident but it should be noted
that it occurs only for a very narrow range of $\lll$ values. Of course, $\lll$
is a parameter that is determined by other variables and is not fixed 
externally. The bulk pressure is no longer given by the large-$a$ expression
(equations (\ref{P_LA}, \ref{P_LA0}) but is well fitted by the small-$a$ result
(equations (\ref{GAMMA0}), \ref{P_SA}). The prediction is $P_{bulk} = 0.243529$ and
the computed value is $P_{bulk} = 0.243531$. This value is shown for reference 
in figure \ref{f9}. 

The curves for the surface charge, $\ss$, are similar to those of previous 
sections and are not reproduced here. The small $L$ behavior is again consistent
with equation (\ref{eq:sigex}).

The anion and cation number densities (equation (\ref{D_MAT}) are plotted for
$\lll=.123$ and $L=0.1, 0.5 1.1$ in figure \ref{f10}. It is interesting to note
that for both species the density falls sharply at the film surface and the
anion density reaches a peak for the thicker films which is located only a 
short distance into the film. The position and shape of this peak is independent
of thickness $L$ and seems to be a universal feature of the low temperature case.
Only the lowest 4 Mathieu modes make an appreciable contribution since the 
effective charge is large and from equation (\ref{D_MAT}) this causes 
a strong exponential suppression on all but the lowest modes (note that in 
equation (\ref{D_MAT}) a factor of $e^2/2$ is absorbed into all lengths).
Also, the values of $\lll$ in the collapse region decrease as $e$ increases and
so the surface function $f(\phi)$ (equation (\ref{F})) oscillates less fast and 
only has appreciable overlap with the lowest modes. For these reasons the 
species number densities are dominated by the contributions from the lowest modes
and so show more structure at low temperature than at high temperature. This is to
be expected since the electrostatic energy dominates the thermal energy.
We have observed similar maxima in the density profiles for other values of $e$ 
if $\lll$ is sufficiently small. This is due again to the dominance of only a very
few low-lying Mathieu modes.

\section{Comparison with Poisson-Boltzmann Theory}

The Poisson-Boltzmann (PB) theory for our system may either be derived
directly by standard thermodynamic techniques \cite{isra,books},
or as the mean field
theory for the field theory (\ref{eq:FT}). The theory has been used in a
wide context in soft condensed matter physics and in particular to
analyze the behavior of soap films in \cite{exkokh, dese1},
and also in the context of colloidal stability \cite{books}.
In general it is fair
to say that it has been reasonably successful in predicting the
physics of systems where interplane distances are reasonably large
and for monovalent ionic species \cite{andel}.

The resulting equations are
(again scaling so that $\beta = 1$)
\beq {d^2 \phi \over dx^2} = 2\mu e \sinh( e \phi) + \lll e
(\delta(x)\exp( e \phi(0)) + \delta(x-L))\exp( e \phi(L))),
\eeq
where here $\phi$ is the mean field electrostatic
field. Assuming symmetry about the point $L/2$ (however see the
comments in the conclusion) and using the condition of
electroneutrality (which is a mean field assumption), 
the boundary conditions are 
\beq {d \phi \over dx}|_{0,L} = {d \phi \over dx}|_{L\over 2} = 0 \eeq 
Interestingly $\phi$ appears as a purely imaginary  saddle point of
the theory (\ref{eq:FT}). In the region $[0^+, L^-]$ the above equation
reduces to 
\beq 
{d^2 \phi \over dx^2} = 2\mu e \sinh( e \phi),\label{eq:PB1}
\eeq 
with the boundary conditions
\beq
 {d \phi \over dx}|_{L\over 2} = 0 \label{eq:PB2}
\eeq
and 
\beq
{d \phi \over dx}|_{0} = -\lll e \exp( e \phi(0)) = \sigma,
\label{eq:PB3}
\eeq
where $\sigma$ is the surface charge. It easy to show that the mean field
free energy over the bulk is \cite{isra} 

\beq 
F_{MF} = \int^L_0(\phi^\prime)^2\,dx~-~\lb 2\lll \exp(\beta e \phi(0)) + 2\mu
L(\cosh(e\phi(L/2)) - 1)\rb~.
\eeq
Then we find 
\beq
 P_d = {2\mu} (\cosh(e\phi_m) - 1)~,
\eeq 
where $\phi_m = \phi(L/2)$ is the midplane potential.
One immediately sees that in the case $\lll = 0$ then $\phi = 0$ is
a solution and the film is always marginally stable in the sense that
$P_d = 0$. In general any non-zero $\lll$ gives a non-zero value of
$\phi_m$ and hence the film is always stable for non-zero
$\lll$. This is clearly at variance with the exact results derived
here. Moreover, the mean field bulk pressure is $P_{bulk} = 2\mu$ which
is only applicable to the limit $e \rightarrow 0$ or, equivalently,
$a \rightarrow \infty$. 

In general one must resort to a numerical solution of the above mean
field equations. However in the case where $L$ is small such that
$\phi$ varies only slightly we may use the approximation,

\beq 
\phi = \phi_m + C (x- {L\over 2})^2.
\eeq

Substituting this into equation (\ref{eq:PB1}) yields $C = \mu e
\sinh(\phi_m)$. Using this in the boundary condition (\ref{eq:PB3})
then yields 

\beq 
\mu L \sinh(e \phi_m) = -\lll \exp(e \phi_m) + O(L^2).
\eeq

Solving this yields

\beq 
\phi_m = -{1\over 2e}\log( {2\lll \over \mu L} + 1).
\eeq

Hence in this limit
\beq 
P_d(L) \sim \sqrt {2\lll \mu \over L} \label{eq:pdmf}
\eeq 
and
\beq 
\sigma \sim - e\sqrt {L\mu\lll \over 2}.
\eeq
One sees that while the surface charge $\sigma$ does decay to
zero it does so as $L^{1\over 2}$ in comparison with the exact result
(\ref{eq:sigex}). In addition the disjoining pressure (\ref{eq:pdmf})
actually diverges rather than tending to a constant. 

For infinitely thick films one may use the condition that $\phi_m \to 0$
in order to calculate the surface potential. In this case the surface
potential is given by the equation

\beq 
\cosh(e\phi(0)) - {1\over 4\mu}\lll^2 e^2\exp(e\phi(0)) = 1 ,
\eeq
from this we find that the physical solution is 
\beq \phi(0) = -{1\over e} \log(1 + \lll e/\sqrt{2 \mu}) \eeq
giving a surface charge
\beq
\sigma = {-\lll e \over 1 + \lll e/\sqrt{2\mu}}.
\eeq 

At intermediate distances one has to numerically solve the PB
equations. For the cases discussed in section \ref{INTER} there
is no agreement at all between the numerical solution to the
PB equation and the exact result. This is to be
expected since there is no collapse predicted by the PB equation.
However, there is no agreement even on the rising part of the pressure
curve at $L$ much greater than that at the pressure maximum. 
Also, the values of $P_{bulk}$ are not close to the mean-field prediction
of $P^{MF}_{bulk} = 2.0$ although for $e=0.1$ this value is 
approached. Nevertheless, in this latter case there is still a large 
disagreement between the exact and mean-field curves. Indeed, 
we have investigated very small values of $e$ for a large range
of $\lll$ values but have not found any reliable agreement 
between the exact theory and the PB equation. The PB equation
may be applicable for even smaller values of $e$ than we have investigated. 
Indeed, a naive analysis of the  applicability of the saddle point
 method for the theory (\ref{eq:FT}) suggests that $a$ should
be large, i.e., either $\mu \gg 1$ or $e^2 \ll 1$, thus giving
either $\mu$ or $1/e^2$ as the large parameter justifying the saddle point
analysis. However, in the cases we have analyzed here, mean-field theory 
and the PB equation are of very little value in the analysis
of the one-dimensional Coulomb gas.

\section{Conclusions}

In conclusion we have derived an exact solution for the one
dimensional Coulomb gas with boundary effects. Surprisingly the mere
presence of a boundary, without any surface adhesion term, leads to a
reduction of the density near the boundary with respect to the
bulk. This effect means that the disjoining pressure of the system is 
negative and the resulting film will tend to collapse. When $\lll > 0$ we
have shown that at sufficiently large distances the disjoining pressure must
be positive, and hence a stable {\em common} film regime exists. However, if 
the value of $\lll$ is not too large a collapse phenomena may occur where 
the disjoining pressure decreases as the surfaces come together. The disjoining
pressure may even become negative signalling the onset of strong attractive 
forces in the system; this may well be the one dimensional version of the
collapse to a NBF seen in experimental systems. We have also seen the 
possibility of secondary collapses in the parameter ranges we have studied; 
it would be interesting if one could find an experimental system 
exhibiting a secondary collapse. In principle multiple collapses are possible,
but we have yet to see more than two. 
 
Poisson-Boltzmann theory predicts a stable film for any non-zero value
of $\lll$ and in addition the calculated mean field disjoining
pressure is larger that that of the exact calculation. Taking into
account the full theory and all its correlations does indeed introduce
an attractive interaction over and above the mean field result, in accordance
with the calculations made in three dimensional systems using
techniques beyond mean field theory \cite{spbe1,spbe2,po}. We would like to 
comment here that in our solution
of the mean field equations we have, as is done throughout the
literature, always assumed that the mean field solution is symmetric
about the midplane of the film. In physical terms this seems quite
plausible for thick films where the two planes do not interact and
hence there can be no breaking of spatial symmetry. The variant of
mean field theory used in \cite{dese1, dese2} uses this symmetric
solution and the theory describes extremely well both surface tension
data for SDS bulk solutions and the disjoining pressure isotherms up
to the point where the collapse occurs. One may show \cite{dehose}
in the field theoretic sense that the mean field solutions we have 
found here and in \cite{dese1,dese2} are indeed stable local minima of the free
energy and hence they at least describe a metastable state. The fact
that the mean field solutions work so well in this context up to the
collapse point suggest that another mean field solution with a broken
spatial symmetry and possibly with a complex part may appear with a
lower free energy than that of the symmetric real solution. This would
also be consistent with the experimental indications (and indeed conclusions
 that may be drawn from our exact solutions in 1 D) that the
transition to a Newton Black Film is of first order. Work on this
problem is currently under progress \cite{dehose}.

\vskip 0.5 truecm
\noindent{\bf Acknowledgments}
\vskip 0.5 truecm
The authors would like to thank J.J. Benattar, 
R. Bidaux, I.T. Drummond and H. Orland for useful discussions.

\newpage

\bef
\vskip 5mm
\bec
\epsfig{file=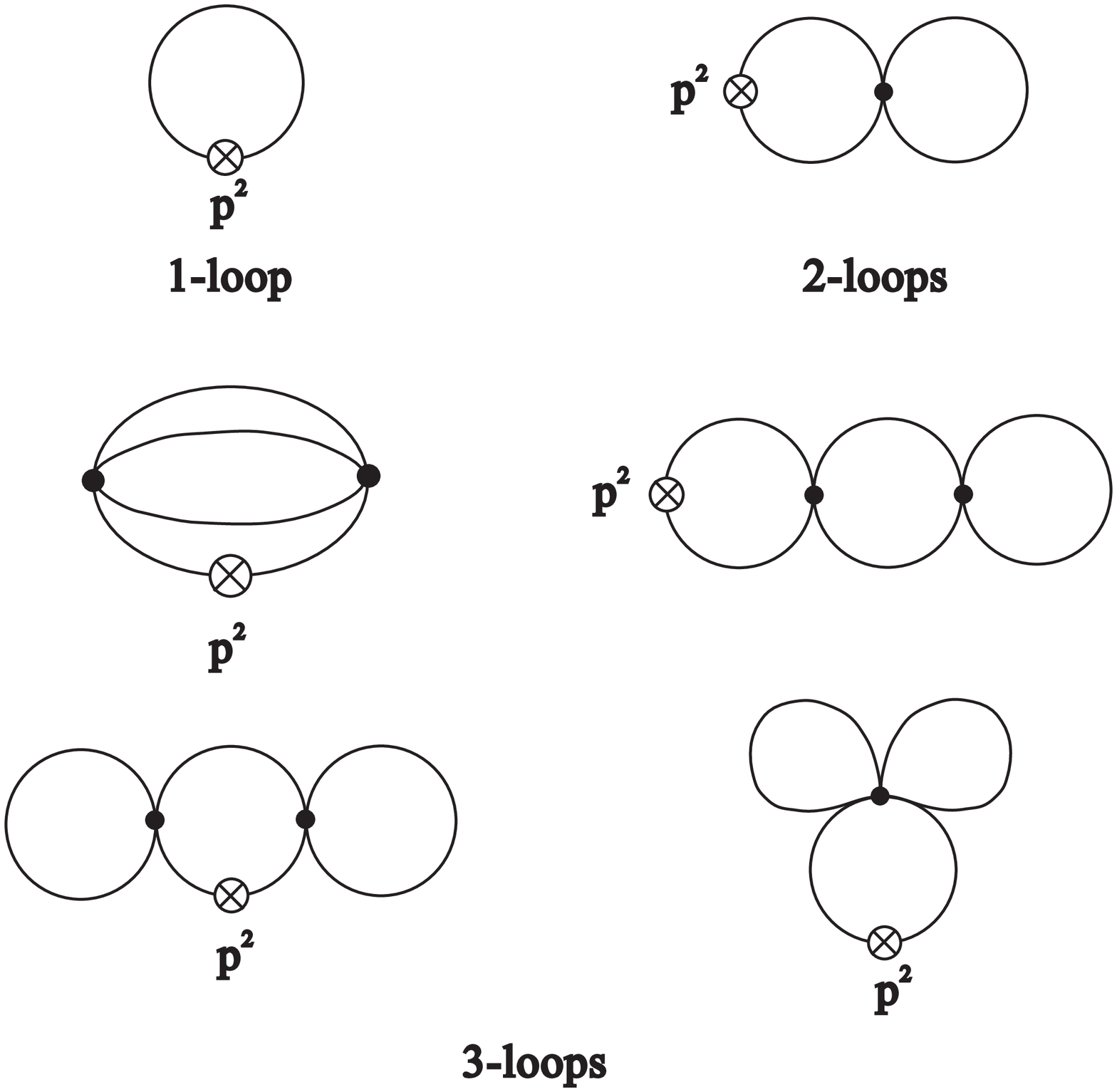,height=80mm}
\enc
\caption[]{\label{f1}
The one-dimensional Feynman graphs up to three-loop order which contribute to
the calculation of the bulk pressure in equation (\ref{P_FD}). The operator
insertion \epsfig{file=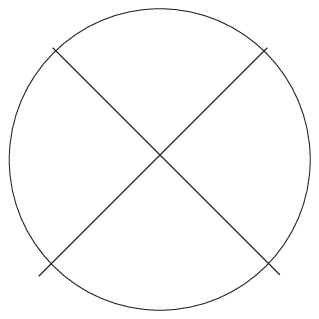,height=3mm} for $\lb\psi^\prime\rb^2$ is shown and
corresponds to the insertion of the factor $p^2~$ in the appropriate loop integral.}
\vskip 5mm
\enf

\newpage
\bef                                                                 
\bec                                                                    
\epsfig{file=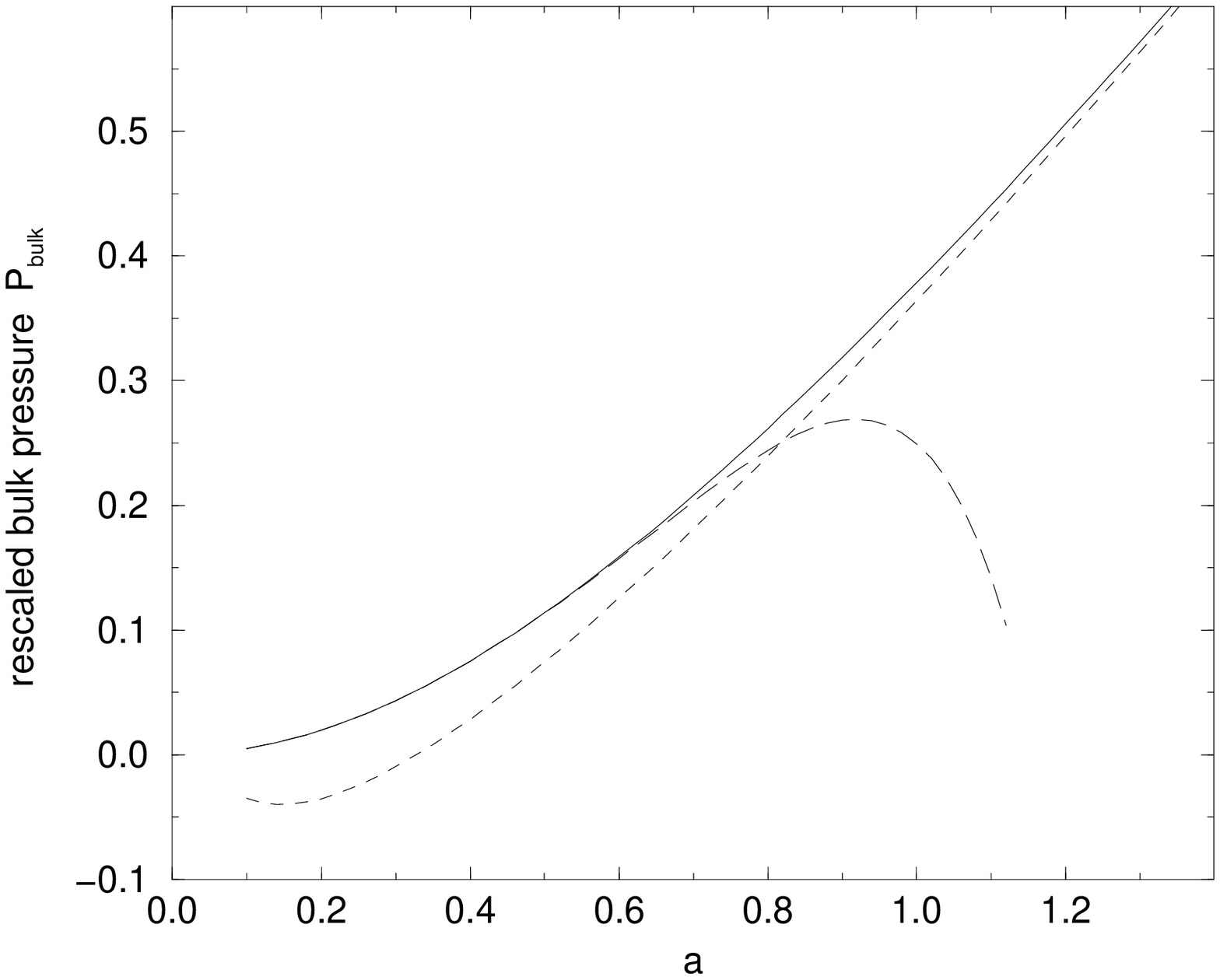,height=120mm}
\enc                                                                   
\caption[]{\label{f11}
The bulk pressure $P_{bulk}$, scaled by $e^2/2$ versus $a = 4\mu/e^2~$. The solid line
is the exact computed curve and the dashed and long-dashed curves are the 
predictions of equations (\ref{GAMMA0}) and (\ref{P_LA}), respectively.
The predictions fit very well except in the region $0.7 < a < 1.1~$.
}
\enf   
\newpage
\bef                          
\bec                       
\epsfig{file=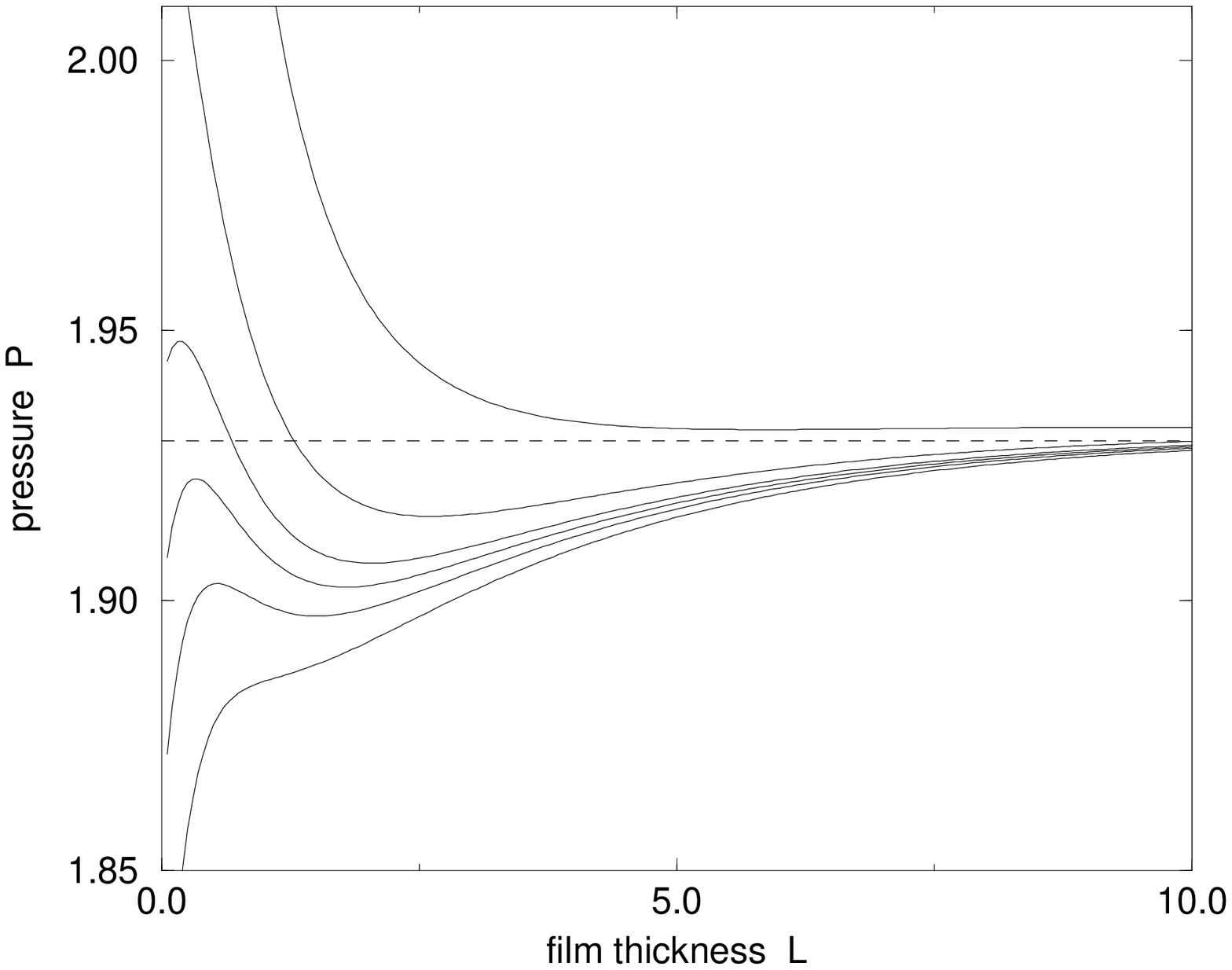,height=120mm}
\enc                                                    
\caption[]{\label{f2}             
The pressure $P$ versus film thickness $L$ for $kT=1.0$, $e=0.1$ and $\mu=1.0$.
for $0 < L \le 10.0$. The different curves are for $\lll = 0.9,0.93,0.95,0.97,1.02,1.2~$ 
which respectively correspond to the curves from lowest to highest pressure at any
given $L$. The phenomenon of primary and, in some cases, secondary collapse are
clearly visible.
}
\enf           
\newpage
\bef
\bec
\epsfig{file=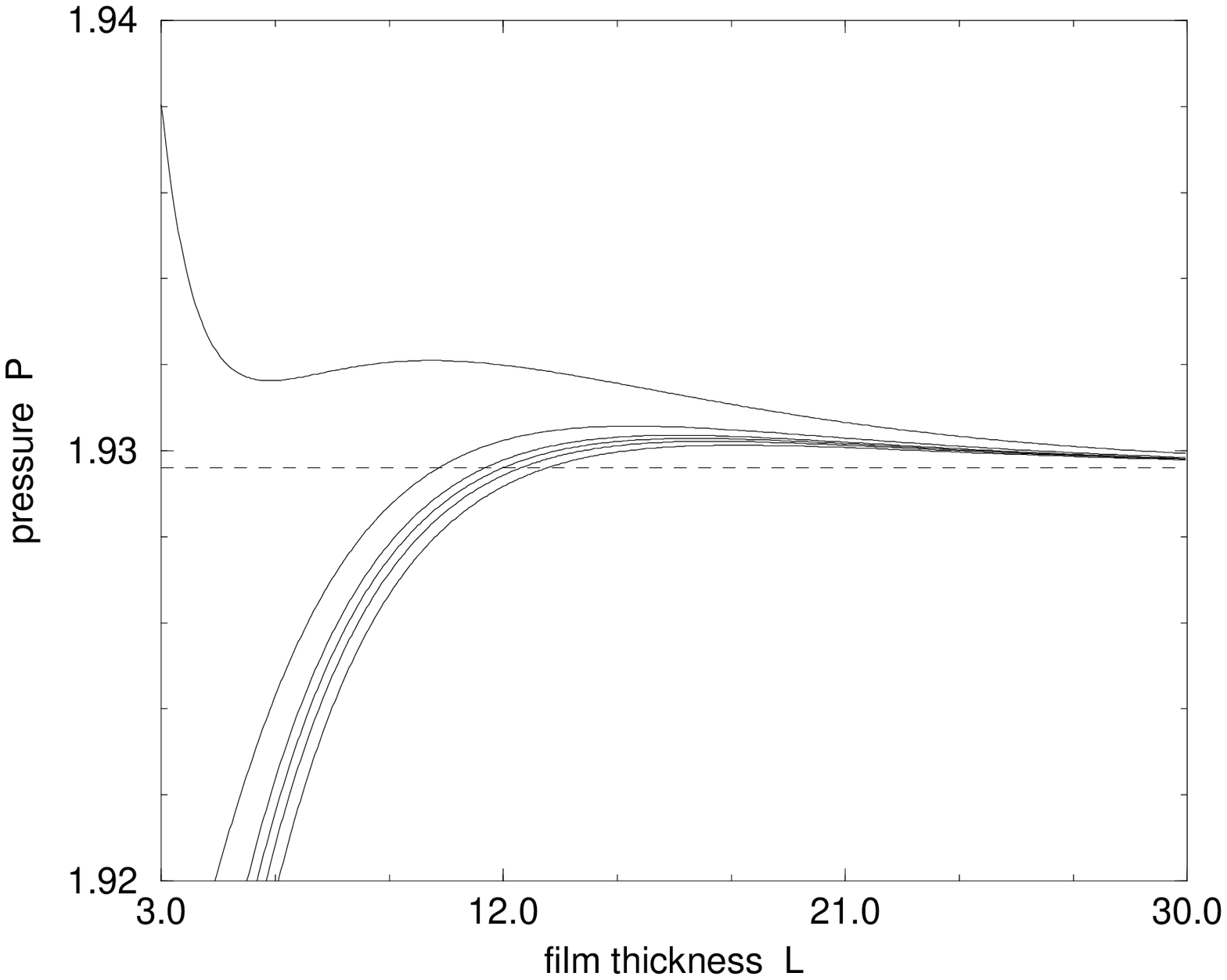,height=120mm}
\enc
\caption[]{\label{f3}
The curves shown in figure \ref{f2} extending to larger $L$ to highlight the region
of primary collapse. As in figure \ref{f2} the higher the value of $\lll$ the
higher the pressure at given $L$.
}
\enf
\newpage
\bef                                                 
\bec                                                       
\epsfig{file=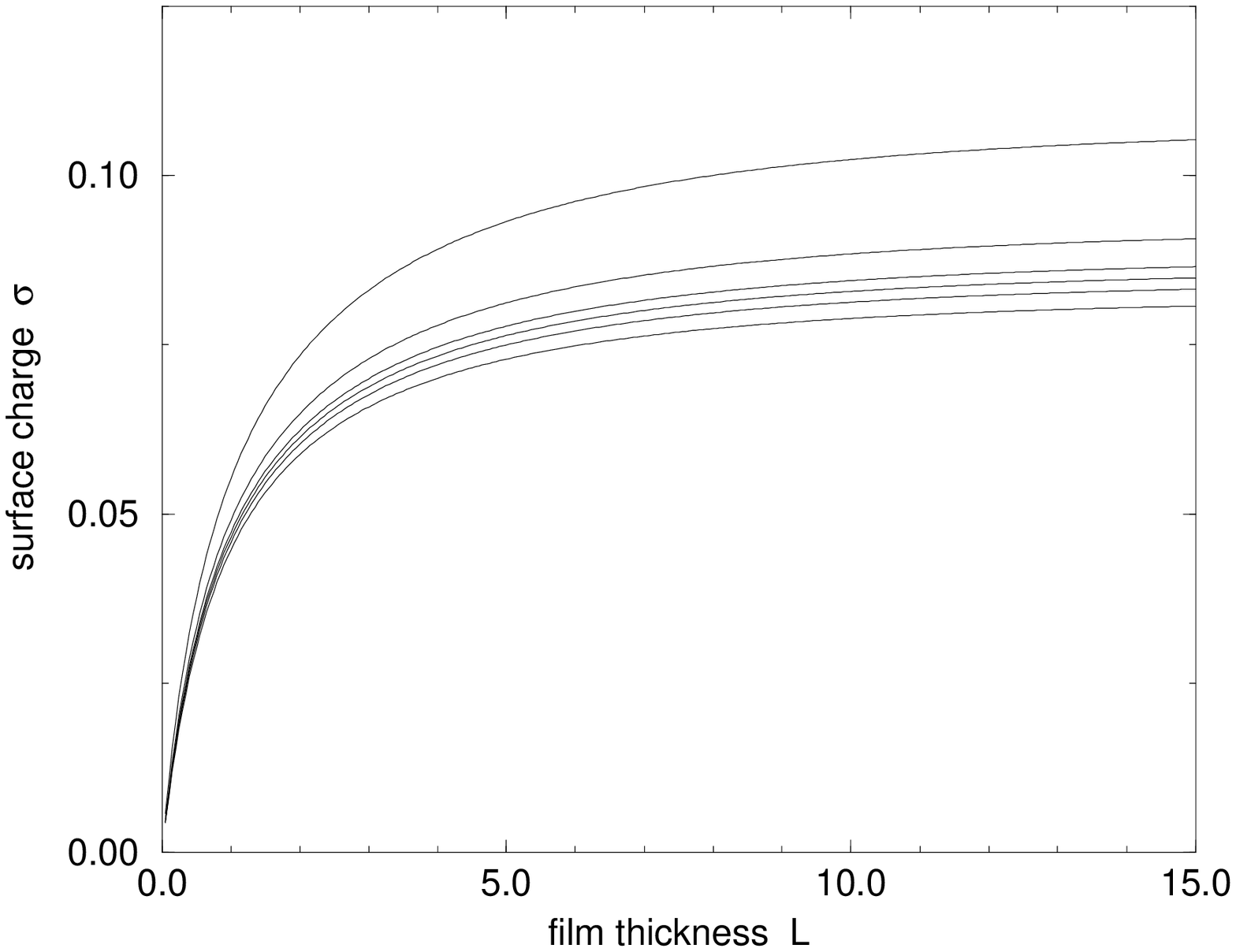,height=120mm}                          
\enc                                                                
\caption[]{\label{f4}                                            
The surface charge $\ss$ defined by equation (\ref{eq:sigex}) versus film thickness $L$
for $kT=1.0$, $e=0.1$ and $\mu=1.0$. The different curves are for 
$\lll = 0.9,0.93,0.95,0.97,1.02,1.2~$ corresponding to the curves from lowest
to highest $\ss$. There is no feature which hints at the presence of the collapse
phenomenon appearing in the pressure curves shown in figures \ref{f2} and \ref{f3}.
}
\enf
\newpage
\bef
\bec
\epsfig{file=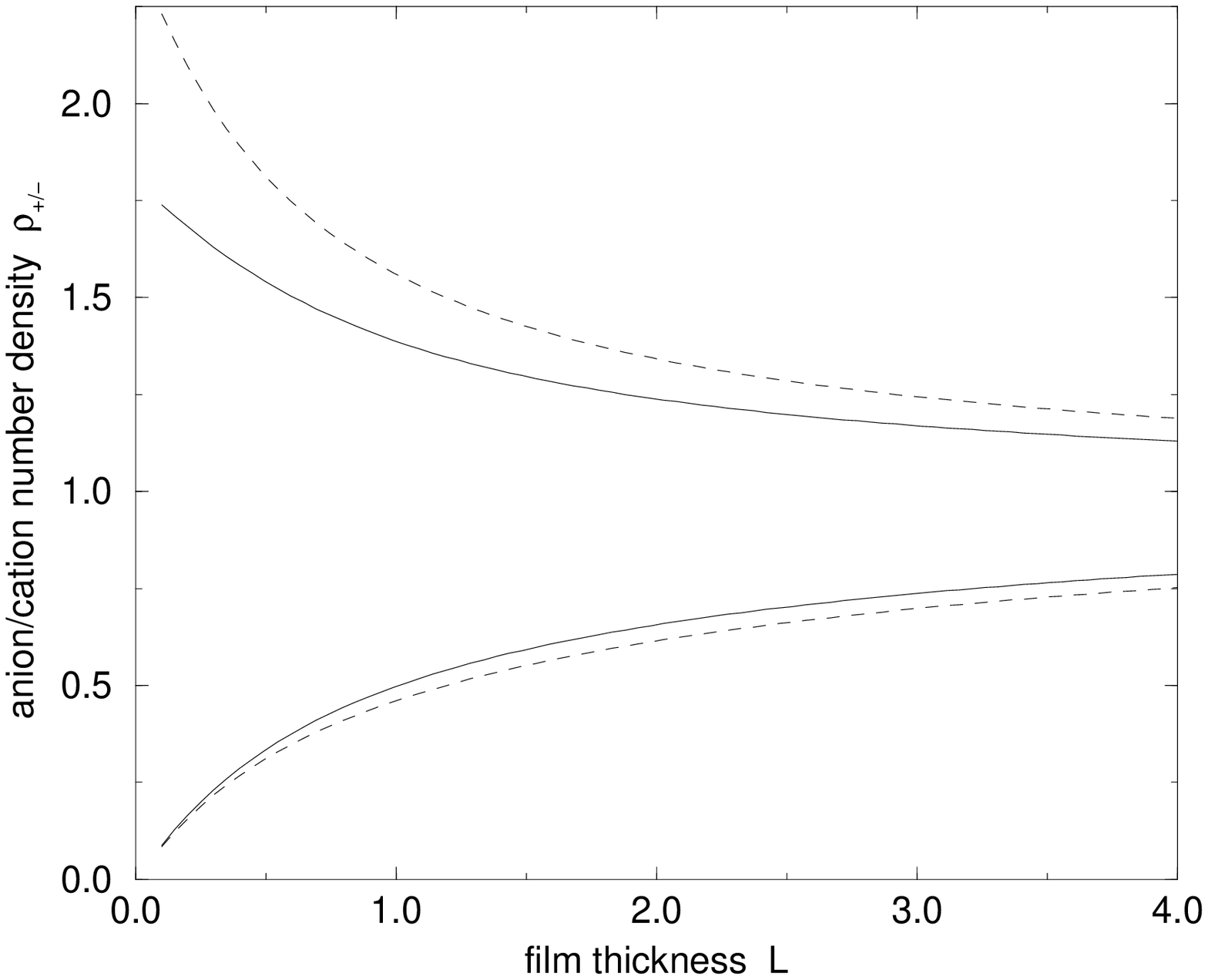,height=120mm}
\enc
\caption[]{\label{f5}
The midplane anion (upper curves) and cation (lower curves) densities as a function of
film thickness $L$, for $e=0.1, \mu=1$, $\lll=0.9$ (solid) and $1.2~$ (dashed).
}
\enf
\bef
\bec
\epsfig{file=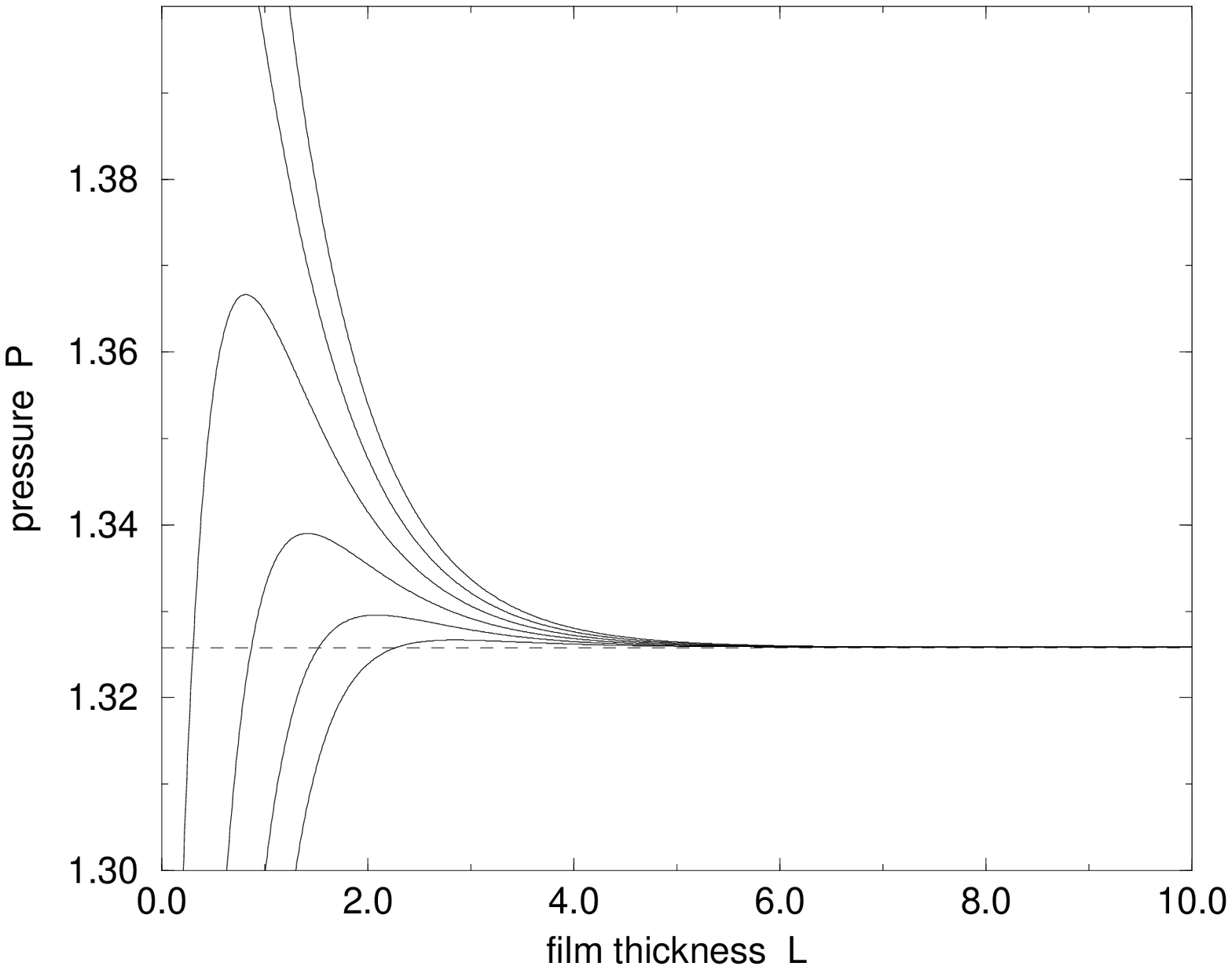,height=120mm}
\enc
\caption[]{\label{f6}
The pressure $P$ versus film thickness $L$ for $kT=1.0$, $e=1.0$ and $\mu=1.0$.
The different curves are for $\lll = 0.3,0.4,0.5,0.6,0.7,0.8~$
which respectively correspond to the curves from lowest to highest pressure at any
given $L$. The collapse phenomenon occurs for the smaller $\lll$ values and disappears
between $\lll=0.7$ and $\lll=0.8$.  
}
\enf
\newpage
\bef
\bec
\epsfig{file=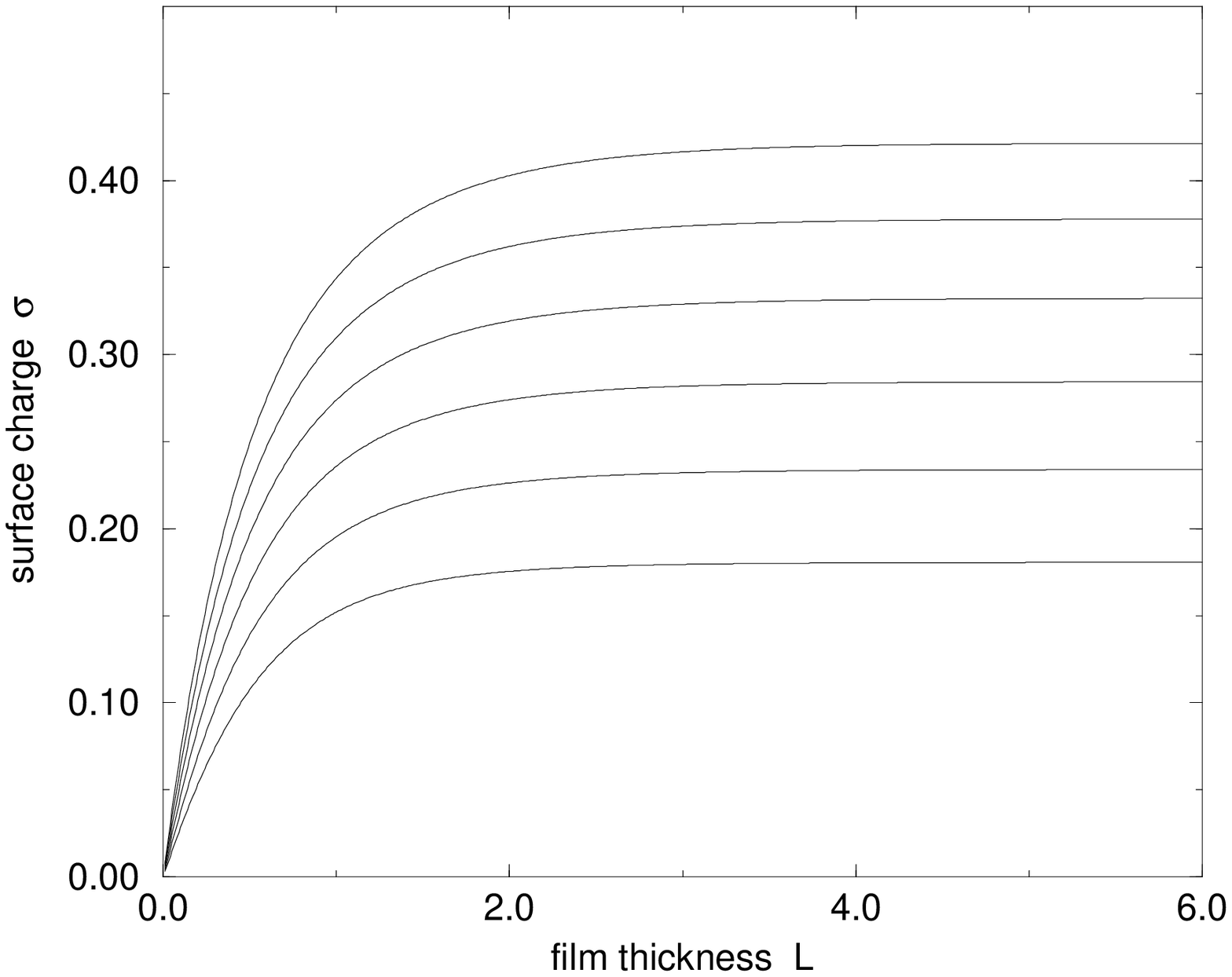,height=120mm}
\enc
\caption[]{\label{f7}
The surface charge $\ss$ defined by equation (\ref{eq:sigex}) versus film thickness $L$
for $kT=1.0$, $e=0.1$ and $\mu=1.0$. The different curves are for
$\lll = 0.9,0.93,0.95,0.97,1.02,1.2~$ corresponding to the curves from lowest
to highest $\ss$. There is no feature which hints at the presence of the collapse
phenomenon appearing in the pressure curves shown in figures \ref{f2} and \ref{f3}.
}
\enf
\newpage
\bef
\bec
\epsfig{file=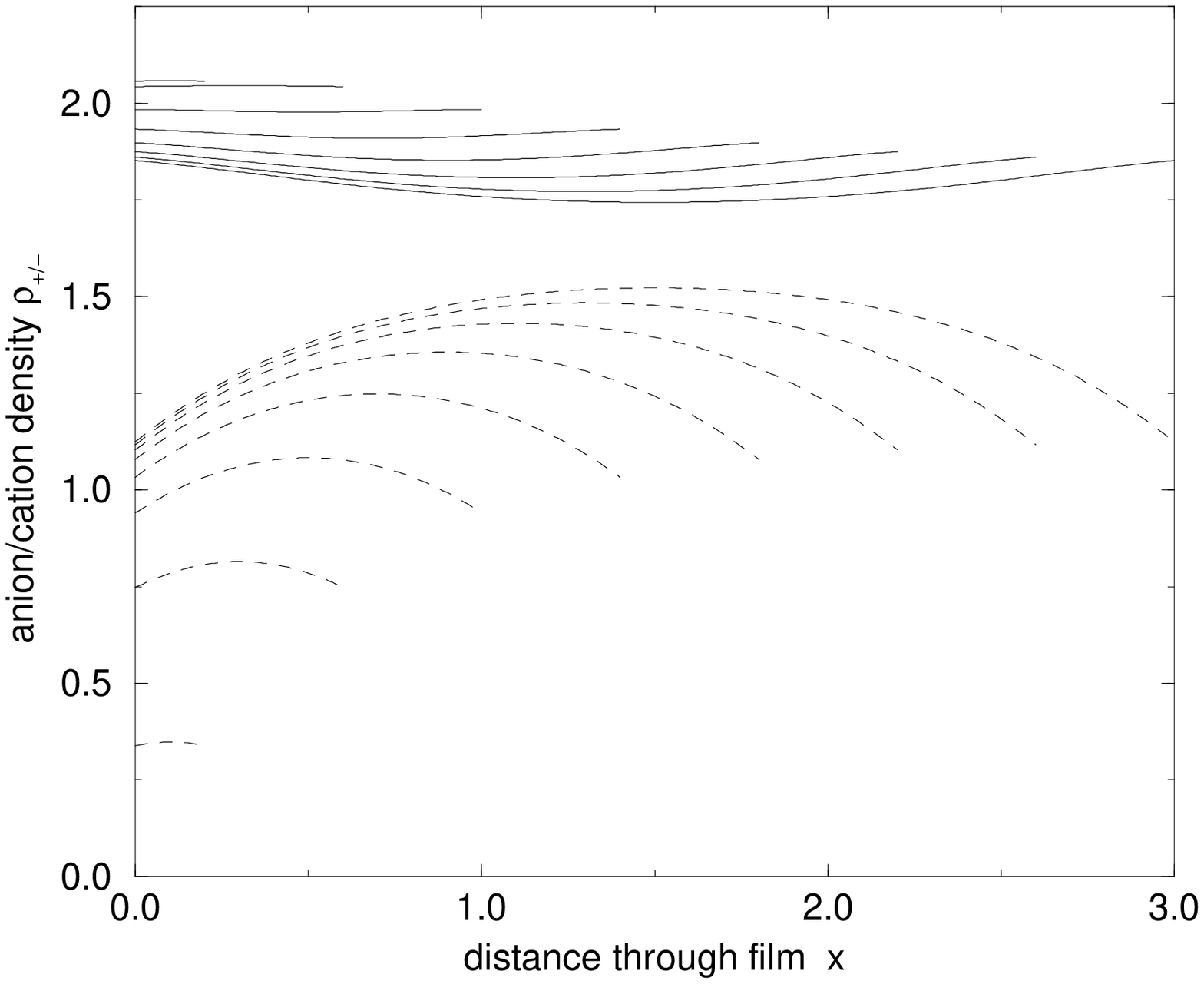,height=120mm}
\enc
\caption[]{\label{f8}
The anion (solid curves) and cation (dashed curves) densities as a function of
distance $x$ through the film for $e=1.0, \mu=1$, $\lll=0.95$ and for various
film thickness $L$. 
}
\enf
\newpage
\bef
\bec
\epsfig{file=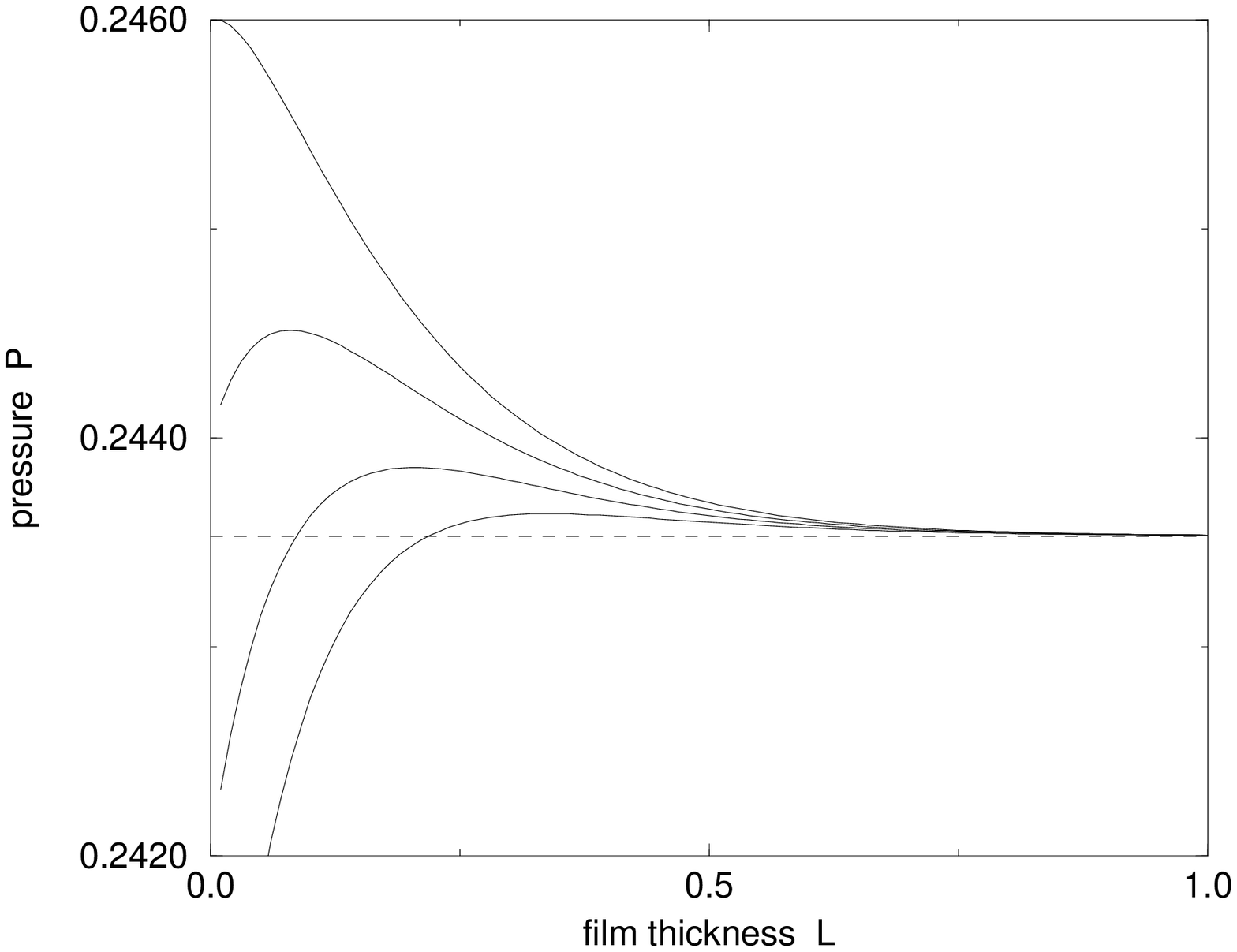,height=120mm}
\enc
\caption[]{\label{f9}
The pressure $P$ versus film thickness $L$ for $kT=1.0$, $e=4.0$ and $\mu=1.0$.
The different curves are for $\lll = 0.12, 0.121, 0.122, 0.123~$
which respectively correspond to the curves from lowest to highest pressure at any
given $L$. The collapse phenomenon occurs for the smaller $\lll$ values and disappears
between $\lll=0.122$ and $\lll=0.123$.
}
\enf
\newpage
\bef
\bec
\epsfig{file=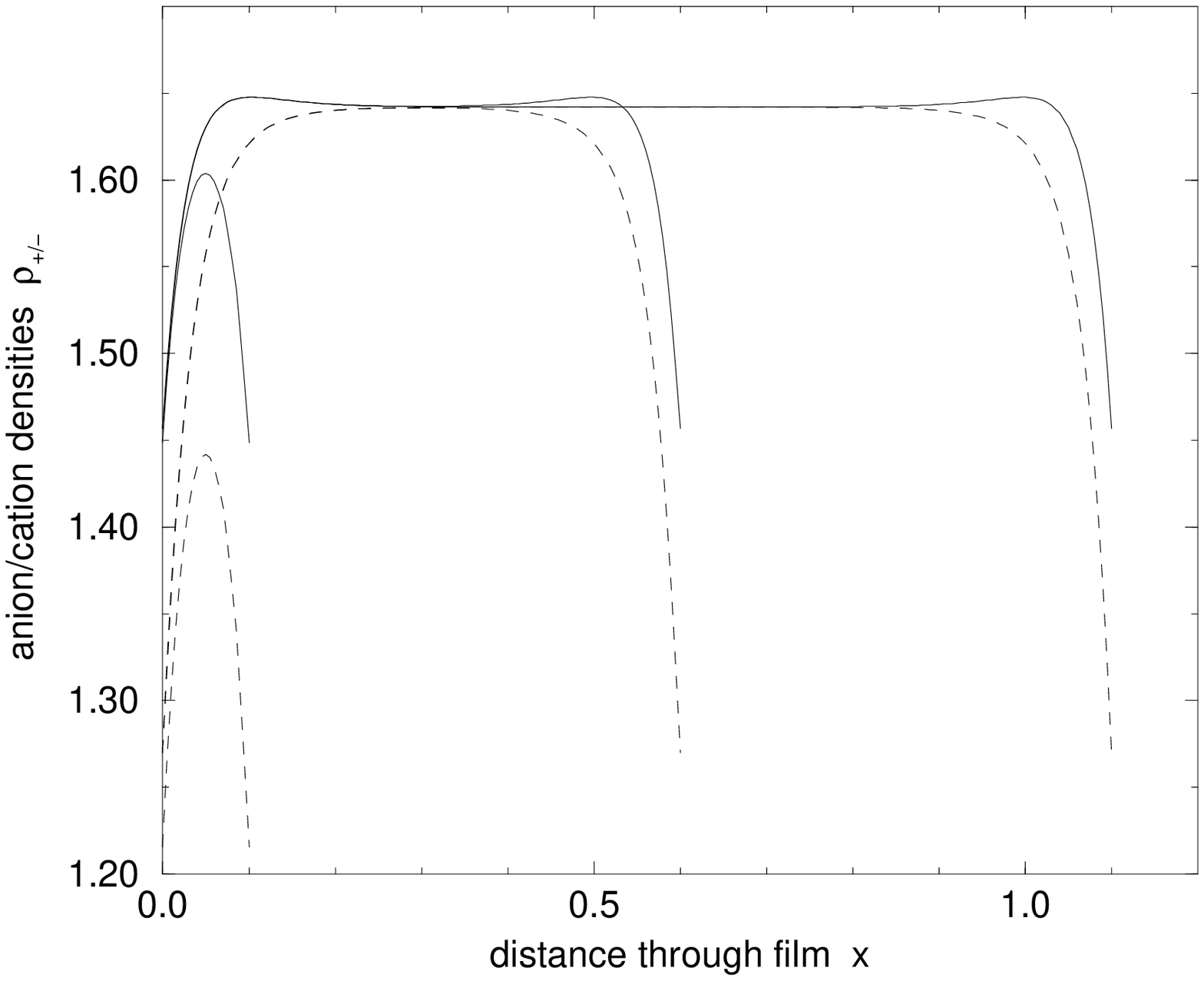,height=120mm}
\enc
\caption[]{\label{f10}
The anion (solid curves) and cation (dashed curves) densities as a function of
distance $x$ through the film for $e=4.0, \mu=1$, $\lll=0.123$ and for film thicknesses 
$L=0.1, 0.5, 1.1$. 
}
\enf

\end{document}